\documentclass[aps,pra,twocolumn,amsfonts,floatfix,superscriptaddress,showpacs,nofootinbib]{revtex4}

\pdfoutput=1

\usepackage{graphicx}
\usepackage{dcolumn}
\usepackage{bm}
\usepackage{mathbbol}
\usepackage{verbatim}
\usepackage{dictsym}
\usepackage{multirow}
\usepackage{ulem}

\usepackage{amssymb}          
\usepackage[latin1]{inputenc} 
\usepackage{amsmath}          
\usepackage{color}
\usepackage{epsfig}
 \usepackage{booktabs, tabularx, multirow, array}

\newcommand{\prk}{\pr(K=k)}
\renewcommand{\prk}{\pr_k}
\newcommand{\prone}{\pr(K=1)}
\renewcommand{\prone}{\pr_1}
\newcommand{\przero}{\pr(K=0)}
\renewcommand{\przero}{\pr_0}
\newcommand{\prn}{\pr(\cn=n)}
\renewcommand{\prn}{\pr_n}
\newcommand{\prnone}{\pr(\cn=1)}
\renewcommand{\prnone}{\pr_1}

\newcommand{\beq}{\begin{equation}}
\newcommand{\eeq}{\end{equation}}
\newcommand{\beqa}{\begin{eqnarray}}
\newcommand{\eeqa}{\end{eqnarray}}
\newcommand{\beqan}{\begin{eqnarray*}}
\newcommand{\eeqan}{\end{eqnarray*}}

\newcommand{\pr}{ \mathbb{P}}
\newcommand{\mut}{\tilde{\mu}}
\newcommand{\cn}{\hat{N}}
\newcommand{\mub}{\bar{\mu}}

\begin{document}

\title{{Asymmetric Architecture for Heralded Single Photon Sources}}

\author{Luca Mazzarella}
\email{mazzarella@dei.unipd.it}
\affiliation{Dipartimento di Ingegneria dell'Informazione,
Universit\`a di Padova, via Gradenigo 6/B, 35131 Padova, Italy}
\author{Francesco Ticozzi}
\affiliation{Dipartimento di Ingegneria dell'Informazione,
Universit\`a di Padova, via Gradenigo 6/B, 35131 Padova, Italy}
\affiliation{\mbox{Department of Physics and Astronomy,
Dartmouth College, 6127 Wilder Laboratory, Hanover, NH 03755, USA}}
\author{Alexander V. Sergienko}
\affiliation{Department of Electrical and Computer Engineering, University of Boston, 8 Saint Mary's St., Boston, MA 02215, USA}
\author{Giuseppe Vallone}
\affiliation{Dipartimento di Ingegneria dell'Informazione,
Universit\`a di Padova, via Gradenigo 6/B, 35131 Padova, Italy}
\author{Paolo Villoresi}
\affiliation{Dipartimento di Ingegneria dell'Informazione,
Universit\`a di Padova, via Gradenigo 6/B, 35131 Padova, Italy}
\date{\today}

\begin{abstract}
{Single photon source represent a fundamental building block for optical implementations of quantum information tasks ranging from basic  tests of quantum physics to quantum communication and high-resolution quantum measurement. In this paper we investigate the performance of a multiplexed system based on asymmetric configuration of multiple heralded single photon sources. {To compare the effectiveness of different designs we introduce a single-photon source performance index that is based on the value of single photon probability required to achieve a guaranteed  signal to noise ratio.} The performance and scalability comparison with both currently existing multiple-source architectures and faint laser configurations reveals an  advantage the proposed scheme  offers in realistic scenarios. This analysis also provides insights on the potential of using such architectures for integrated implementation.}

\end{abstract}

 \maketitle

\section{Introduction}

The ideal source of single-photon quantum states is a key instrument for successful implementation of many exciting quantum information topics ranging from the schemes to probe foundations of quantum mechanics to super-resolution measurement and quantum metrology.
Single photon sources (SPSs) represent also a key resource for optical quantum computing and quantum communication. 
Optical quantum computers  based on integrated {photonic} technology \cite{poli08sci, matt09npho,sans10prl,cres11nco}
can be build  using linear optics and SPSs as shown by Knill, Laflamme and Milburn \cite{knil01nat}. 
In reality,  specific designs that offer only some approximation of an ideal source can be achieved.
For example, the current Quantum Key Distribution {(QKD)} systems use weak laser pulses in place of single-photon sources 
\cite{scarani-rev,isabel,bacco,vill08njp,capr12prl} and 
a decoy state technique \cite{hwan03prl,ma05pra} to avoid the photon splitting number (PNS) attack \cite{bras00prl} on the pulses containing more than one photon.
The development of a scheme for producing true single photon states would guarantee that all pulses contain one and only one photon thus allowing  to increase the key generation rate and to avoid any PNS attack. This challenging task has generated extensive efforts that lead to an appearance of multiple designs  of heralded sources of single-photon states.

The single photon source usually used in quantum information applications consists of a faint laser (FL), namely an attenuated and pulsed laser source\footnote{The use of pulsed driving electric fields allows to limit the temporal interval when the photons are expected to exist thus reducing the impact of detector dark counts. }. 
For a coherent source the number of photons in each pulse can be modeled  by a Poisson random variable.
The probability of having $k$ photon in each pulse is given by
\begin{equation}
\label{posisson}
\prk=\frac{\mu^k }{k!}e^{-\mu},
\end{equation}
where $\mu$ is the mean number of photons in each pulse that depends on the power of the laser. 
Two indexes are usually employed in order to evaluate the output quality of the source: 
the {\itshape one-photon probability} {$\prone$} and the {\itshape signal to noise ratio}: 
\begin{equation}
\label{snrf}
\mbox{SNR}:=\frac{\prone}{1-\przero-\prone}={\frac{\mu }{e^{\mu}-1-\mu}}.
\end{equation}
The SNR is the ratio between the one-photon probability and the probability of having more than one photon in the output of the system. This index quantifies a critical quantity of the source: 
the number of multiple photons per pulse.  In optical quantum computing this lead to errors whose effects are hard to detect and correct while in QKD it opens a
possibility for PNS attacks. 
The main limitations of the FL source stems from the fact that $\mu$ is the only tunable parameter. 
This induces a trade off: the value of $\mu$ that maximizes $\prone$ is given by $\mu=1$ and 
corresponds to a value of the one photon probability of $e^{-1}\simeq 0.37$. However, for $\mu=1$ the SNR is equal to $(e-2)^{-1}\simeq 1.39$, a value  that is typically unacceptable for applications requiring a single photon source. Since the SNR is unbounded for $\mu$ approaching zero, 
the mean photon number is usually kept sufficiently low in order to avoid multiple photons events, thus reducing  also the overall probability of single photon emissions.

Several types of architectures with multiple heralded SPS have been proposed in order to overcome such natural limitations of the FL source.
Photons are often generated in such devices by means of spontaneous parametric down conversion (SPDC). In this nonlinear process  an intense laser
pump impinging on a nonlinear crystal leads to probabilistic emission of entangled pairs of photons {(usually called signal and idler)} into
two different spatial modes, with their rate depending on the pump intensity, the nonlinear coefficient value and on the length of the crystal.
It is then possible to ``herald'' the presence of a photon in the signal mode by detecting the correlated twin photon in the idler mode.
A heralded single-photon source based on SPDC  with an overall heralding efficiency  of 83\% has been demonstrated very recently \cite{rame12qph}.

The  scheme based on multiple heralded SPSs  combined with the use of  post-selection has been  originally proposed by Migdall \cite{PhysRevA.66.053805}  in order to enhance the probability of obtaining a single heralded photon. This implementation requires to use $m$-to-$1$ global switch. 
In the same work, the performance of the proposed scheme has been studied considering the finite detection efficiency. 
However, as pointed out in \cite{Shapiro:07}, an efficient implementation of such device is not currently available, and it would be hardly scalable. 
To overcome these problems a symmetric scheme, employing a total of $m-1$ binary
 polarization-switching photon routers arranged in a modular tree structure has been proposed by 
 Shapiro and Wong in \cite{Shapiro:07}. They
 also considered the probability of emitting $n$ photons taking into account the imperfectness or real detectors and optical switches.
An experimental implementation of the scheme along with an essential discussion of its scalability has been pursued in \cite{PhysRevA.83.043814} using $4$ crystals.
Recently, the analysis of multimode emission in SPDC used for SPS was carried out \cite{chri12pra}.

However, despite the recent theoretical and experimental improvement of single photon sources, 
a thorough analysis of the performance of multiple heralded SPS in the presence of finite efficiencies, and their comparison with respect to a simple faint-laser source, is lacking: nonetheless, it appears to be a crucial step in assessing their potential, especially in the light of the experimental difficulties reported in \cite{PhysRevA.83.043814}. 
In pursuing this analysis, we believe that one of the most delicate point consists in devising 
proper performance indexes, ensuring a fair comparisons between different methods. 

In this paper, after reviewing the main ideas and theoretical results underlying the existing SPS architectures, 
we  introduce a proper performance index, given by the single
photon probability for a guaranteed value of the signal to noise ratio, in order to analyze and compare different single photon sources.
We then propose a new and more efficeint heralding scheme based on
an asymmetric configuration.
We finally develop a comprehensive comparison between the architectures that concentrates on the comparative performance, 
the scalability, and the limits of SPS with multiple heralded sources in realistic scenarios.

\section{Performance index for Single Photon Sources}
In this section we present the key ideas underlying   a class of source architectures that outperform the FL scheme
and introduce a performance index for comparing different single photon sources.
 
\subsection{Multiple sources and\\ the advantage of post-selection}

 The building block of such architectures is represented by the so-called Heralded Source (HS). The HS exploits  the SPDC effect on a non-linear crystal pumped with a strong coherent field, which leads (with a certain probability) to the simultaneous generation of 
{a pair (or more pairs) of photons.}: if the duration of the pulse $(\Delta t_p)$ is much shorter than the measurement time interval $(\Delta T)$, but much greater than the reciprocal of the phase-matching bandwidth i.e., $(\Delta \omega)$, i.e. $\Delta T >>\Delta t_p>> 1/\Delta \omega$,
the statistics of the pairs is still Poissonian \cite{Shapiro:07}. 
One photon (the idler) of the pair is then fed to a photon detector, while the other photon is used as signal.
The HS can be employed in  multiple crystal strategies that outperform the FL by exploiting the parallel use of HS units and post selection strategies:  intuitively, the advantage of using a scheme exploiting a parallel implementation lies in the fact that the intensity of the pump of each crystal can be kept low, suppressing thus the multi-photons events, while keeping an acceptable production rate of single photons.

Let us assume for now to employ ideal detectors in order to illustrate in a simple setting the potential advantages offered by this system. The case of finite efficiency will be discussed later in this section. We here consider standard single photon APD detectors
only able to discriminate between the case of no incident photons and the case in which photons are detected, without resolving their number. When a detector is hit by the photons, it returns an electric signal ({\itshape trigger}), which indicates the presence of at least a photon in the signal channel.

\begin{figure}[t]
\centering\includegraphics[width=8cm]{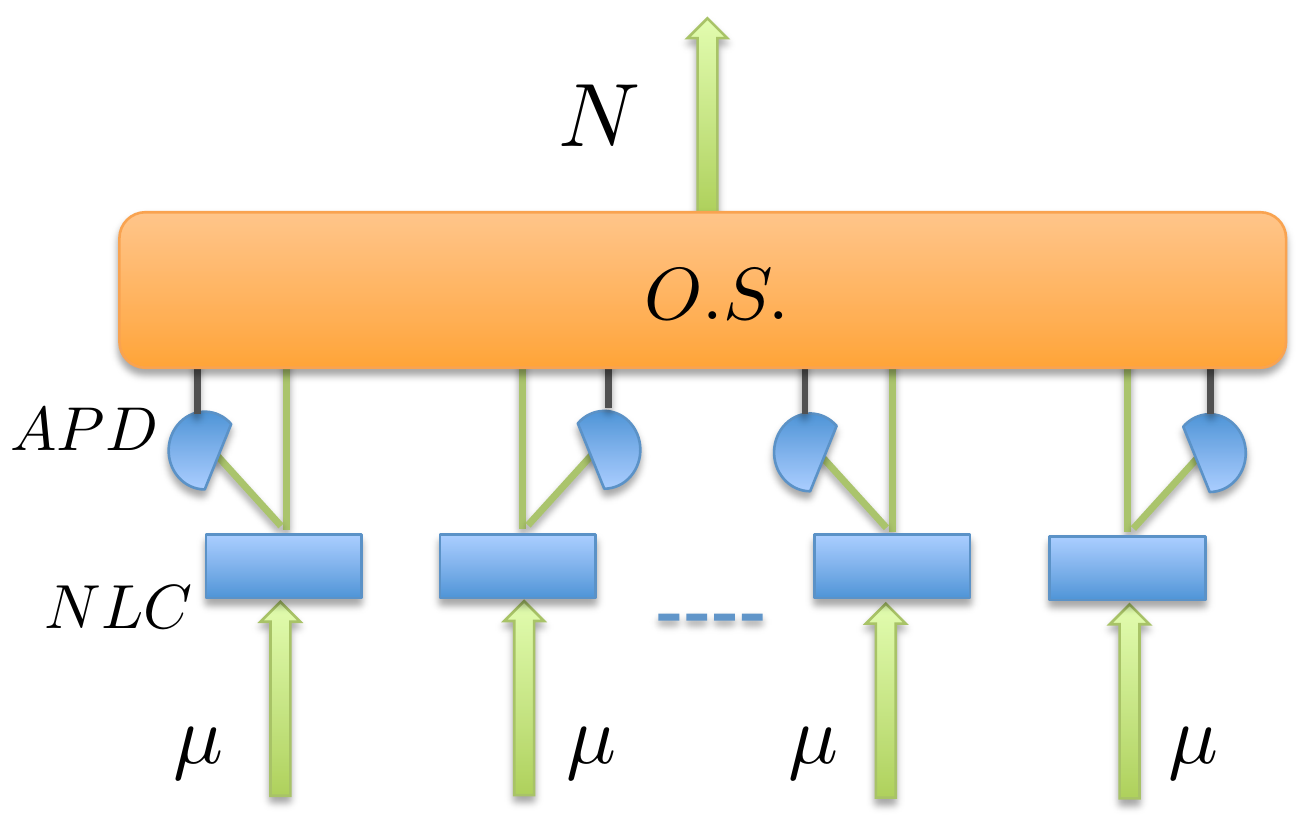}
\caption{Schematic of the MHPS scheme. The blue rectangles labeled with $NLC$ represent the non-linear crystals, the detectors are labelled  with $APD$  and the orange rectangle, labelled with $O.S.$, represents the optical switch. }
\label{idealsys}
\end{figure}
Let us consider a Multiple Crystals Heralded Source Architecture with Post-Selection (MHPS) built as follows:
an array of $m$ HS units, labelled with index  $i=1,\dots,m,$  and each one simultaneously fed with a laser pulse with intensity 
such that the mean number of produced pairs is $\tilde{\mu}$. 
For each HS unit, the idler photon is used as trigger and the other is injected into an optical switch. 
The optical switch selects, depending on the triggers, which 
signal channel must be routed to the global output.

As first proposed in \cite{PhysRevA.66.053805}
we use the following strategy for the switch: the output signal is taken from the first source 
(starting from $i=1$)
that triggers (thus indicating the presence of at least a photon in the channel).
If all the detectors do no fire there are no photons in the signal channel. 
{It is worth noting that the precise structure of the switch is not important in this ideal situation. 
In fact, any switch that selects a channel when at least the corresponding HS has triggered 
could be used without altering the performances.
We will see in the next section how different selection rules affects crucially the performances 
once non-ideal situation are kept into account.}

The probability of having $n$ photon in the global output of the MHPS is given by:
 \beq
\label{MHSn}
  \begin{aligned}
\prn
=&\frac{\mut^n }{n!}e^{-\mut} \frac{1-e^{-m \mut}}{1-e^{- \mut}}(1-\delta_{n})+\delta_{n}e^{-m \mut}.
\end{aligned}
\eeq
In particular, the single photon emission probability and the SNR are given by:
\begin{equation}
\label{snrid}
\prnone={\widetilde \mu}e^{-\widetilde \mu}\frac{1-e^{-m \widetilde \mu}}{1-e^{-\widetilde \mu}},	
\quad
\text{SNR}=\frac{\mut }{e^{\mut}-1-\mut}\,.
\end{equation}
Notice that, when $\mut=\mu$, the signal to noise ratio  is equal to the SNR of the Faint Laser,
while the single photon probability $\prnone$ is always larger.

\subsection{Proposed performance index}
\label{performance}
\begin{figure}
\begin{center}
\includegraphics[width=8cm]{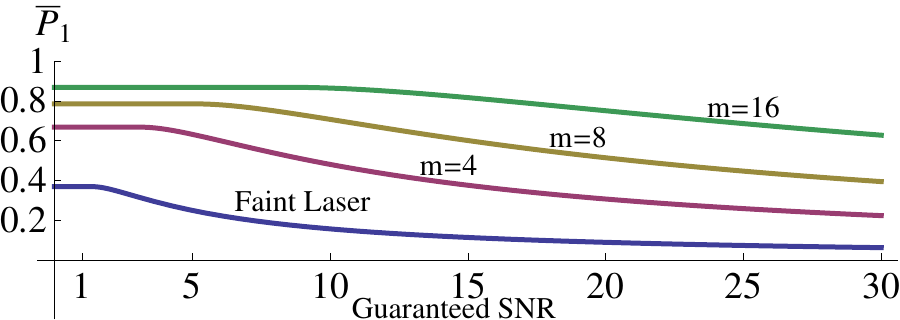}
\caption{ One photon probability $\overline\pr_1$ for the faint laser and the MHPS scheme, 
with $m=4,8$ and $16$ in function of the guaranteed SNR.}
\label{conf}
\end{center}
\end{figure}

As we have stated previously, in many application it is crucial to be able to rely on a threshold value for the SNR. 
With this in mind, we propose the following method to compare different single photon sources: 
we fix a threshold value for the acceptable SNR, 
$\Theta$, and by varying $\mu$
we compute the maximum of the one photon probability ${\pr}_1$ provided that the SNR has a greater or equal value than $ \Theta$,
namely:
\begin{equation}
\overline{\pr}_1( \Theta)
=\max_{\mu,\ \text{SNR}\geqslant \Theta}\prnone(\mu).
\end{equation}
From now on, $\overline{\pr}_1$ always indicates this optimized probability with the SNR constraint.
We note that, since the value of the SNR in eq. \eqref{snrid} is between 0 and $+\infty$, 
by choosing the appropriate value of $\mu$, any value of the SNR can be achieved.
In figure \ref{conf} we show the maximized one photon probabilities of different MHPS schemes
in function of the guaranteed SNR.

 The benefits of the MHPS with respect to the FL are   apparent: fixed any SNR level, it is possible to obtain a  higher value of the one photon probability with MHPS. This is because of the post selection procedure, that can turn (with a certain probability) events in which more than one detectors trigger at the same time into an event that corresponds to a one-photon output {\it by blocking the output of all the HS units but one}.

\subsection{Finite efficiency and symmetric modular architecture} 
Any discussion regarding the physical implementations would be vain without taking into account the realistic efficiencies in detection and routing of the produced photons. An actual detector, in fact, is subject to  losses whose effects are usually described by introducing a parameter $0\leq\eta\leq1$, called {\itshape detection efficiency}, that represents the probability of detection of an incident photon.
This parameter takes into account also the collection efficiency,
as the phase matching relations yield  uncertainty in the direction of the emitted photons.

 Physical implementation of post-selection rules are subject to losses as well. The efficiency in transmission is modeled with a parameter  $0\leq \gamma \leq 1$, called {\itshape transmissivity}, that represents the probability of transmission for a photon through the router. It is important to note that this limited efficiency is referred only to the transmission of the photons: for what it concerns the transmission of the electrical signal   we are always going to assume that, once a trigger happen, it is transmitted until the end of the transmission chain without errors.

 We remark that, in order to consider the role of a finite transmissivity, it is key to specify the particular routing/switching architecture that is being considered, since the potential gain with respect to a FL (without the routing inefficiencies) will in general depend on it. 
 This is not the case in the ideal scheme described in the previous section.

\begin{figure}
\begin{center}
\includegraphics[width=8cm]{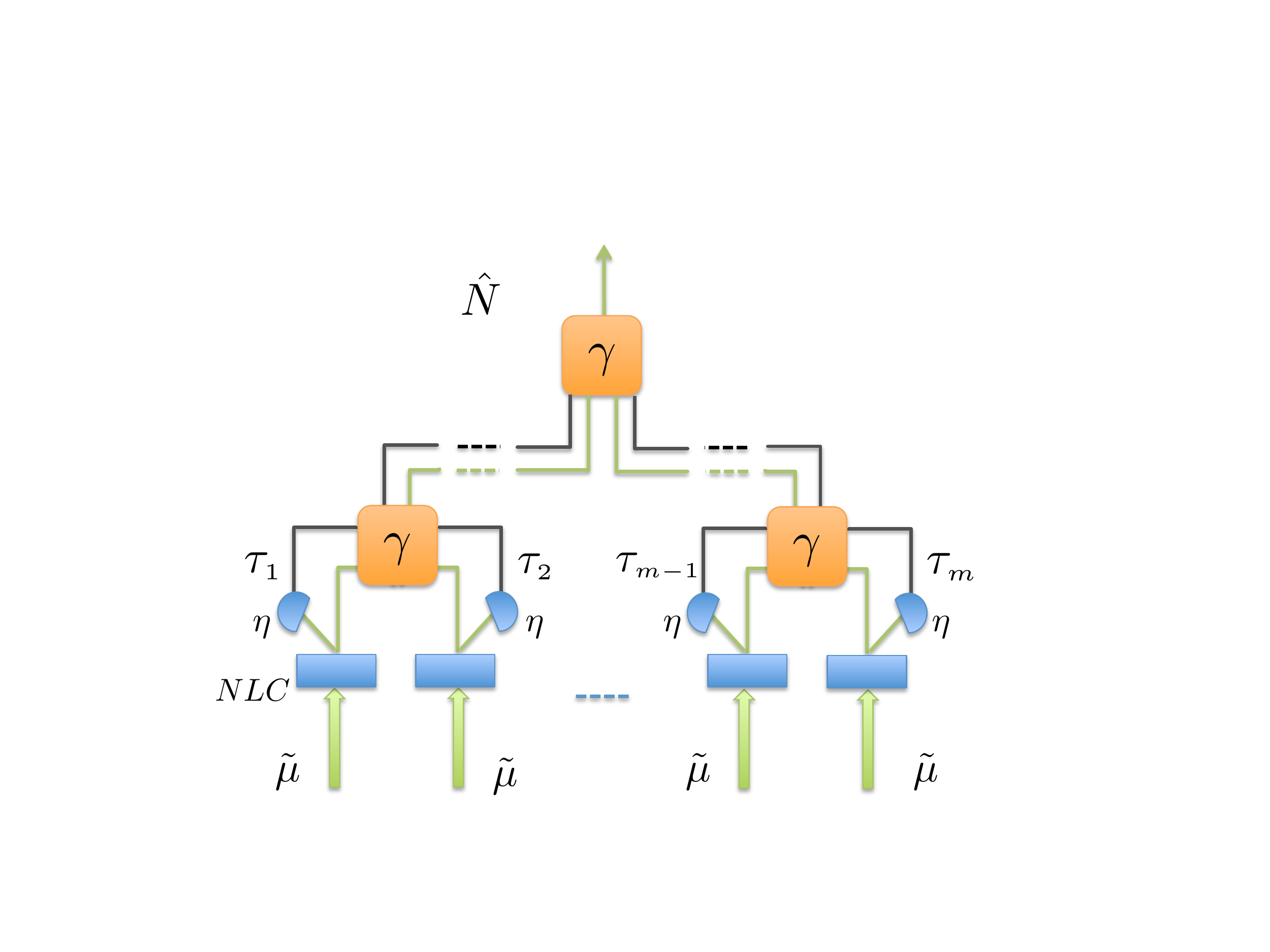}
\caption{Schematic of the SMHPS scheme proposed in \cite{Shapiro:07}. The blue rectangles labeled with $NLC$ represent the non-linear crystals,  the detectors are labelled with $\eta$ and the orange squares, labelled with $\gamma$, are the photon routers.
$\tau_i$ are the trigger signals. }
\label{symm}
\end{center}
\end{figure}
In what follows we are going to briefly review a modular architecture, proposed in \cite{Shapiro:07}, that employs  binary photon routers (2-to-1).
The  $m$-HS units are arranged as shown in  fig. $\ref{symm}$:
the outputs of the first stage are fed into the second  stage's routers   and so on, until the end of the transmission chain. This architecture can be clearly  realized only for  $m=2^k$,  with $k \in \mathbb{N}$.
 It is worth noting that any successfully transmitted photons have to pass trough $k=\log_2 m$ routers.  We will call this scheme Symmetric  Multiple-crystals Heralded-Source   with  Post-selection (SMHPS). 
Each binary router selects the right signal channel only when the
left HS has not triggered and the right HS has triggered: in all other cases it selects the left signal channel.
The overall effect of the routers is that, if more than one detectors fire, the channel routed to the end of the   chain is  the one coming from the crystal corresponding to the lowest $i$. Differently from the scheme proposed in \cite{Shapiro:07}, 
if no detectors fire, the first channel is routed to the end:
in fact, even in this case, there is some probability, due to detection inefficiency, that a photon is generated and
it is convenient to route one channel to the end. With this choice the SMHPS always outperforms the faint laser.

Let's consider that each HS produces a mean number of pair given by $\widetilde\mu/\gamma^k$:
we use this convention to compensate the $\gamma^k$ factor arising from the binary switch transmission.
As shown in Appendix \ref{sec:SMHPScalculation}, the probability of having $n$ photons in the final output is:
 \begin{align}
\label{symprob}
\prn^{S}=&\frac{[(1-\eta )\widetilde\mu]^ne^{-(1-\eta )\widetilde\mu }}{n!}e^{-\eta\widetilde\mu \frac{2^k}{\gamma^k }  }+
\\
&\notag\frac{\widetilde\mu  ^ne^{-\widetilde\mu  }}{n!}\frac{1-(1-\eta )^ne^{-\eta \left(\frac{1}{\gamma ^k}-1\right)\widetilde\mu }}{1-e^{-\eta  \frac{\widetilde\mu }{\gamma ^k}}}(1-e^{-\eta \widetilde\mu \frac{2^k}{\gamma^k }})
\end{align}
The first term in the previous sum accounts for the probability of having some photons in the output when no detector fired.
We notice that, in the ideal case of $\eta=\gamma=1$ we obtain equation \eqref{MHSn}, while in the limiting case of 
 null detection efficiency $\eta$, 
we obtain the faint laser source with mean photon number $\widetilde\mu$: the latter property is related to
the choice of routing the first channel to the global output when no detectors fire.
We will postpone the performances comparison of SMHPS with the FL in section \ref{sec:comparison}.

\section{Proposed asymmetric architecture}
\label{sec:asymmetric}

In the all the previous works, 
it was assumed that all the crystals in the symmetric architecture were driven with same intensity. In Appendix \ref{sec:sym_not_opt} we will 
prove that this symmetric choice represents a suboptimal case for the one photon probability.
In fact, even if the architecture is symmetric, an asymmetry comes from the binary switcher:
the left source is initially checked and, only if this source doesn't trigger, the switch consider the right source. 
This asymmetry can be turned in a resource to increase the one-photon output probability: 
we here propose an asymmetric scheme which is scalable in the number of crystals and that performs better than the 
SMHPS in many situation of experimental interest, still being sub-optimal in exploiting the available resources.

\begin{figure}
\begin{center}
\includegraphics[width=8cm]{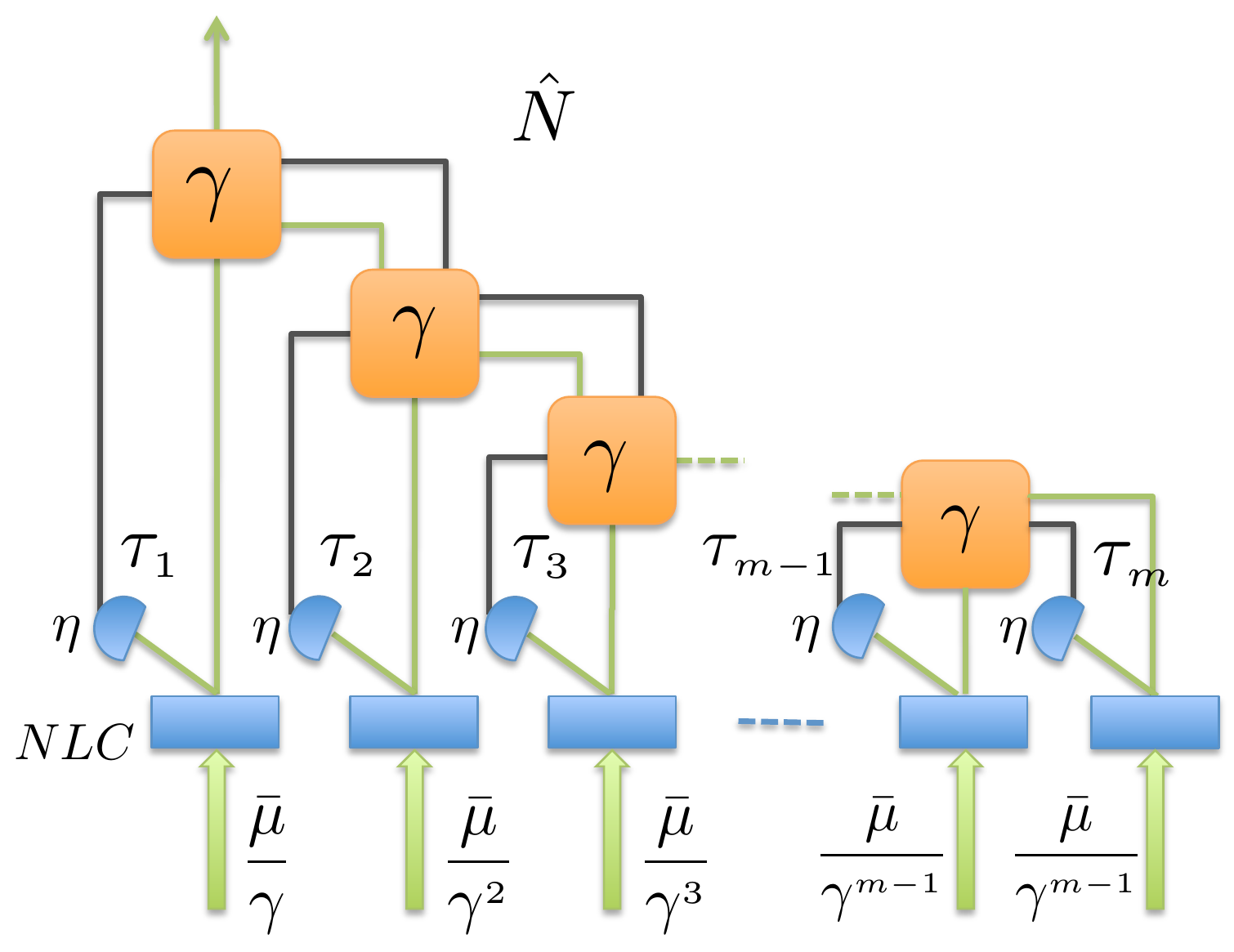}
\caption{Schematics of the  AMHPS. Notice that each crystal is fed with different intensities  to compensate the different absorption rate of different channels.The blue rectangles labeled with $NLC$ represent the non-linear crystals,  the detectors are labelled with $\eta$ and the orange squares, labelled with $\gamma$, are the photon routers. $\tau_i$ are the trigger signals.}
\label{oursystem}
\end{center}
\end{figure}

Let us suppose to have an array of $m$-HS system arranged asymmetrically as fig. \ref{oursystem}. 
This  scheme also employes the same kind of binary switches 
but the multiplexing is performed in a different way with respect to the symmetric configuration: 
the output of a $m$-source block is obtained by combining the output of a block with $m$-1 sources
with the output of a $m$-th source.
An evident advantage with respect to the symmetric configuration is the possibility of adding a single HS without the
constraint of having $2^k$ crystals.

In this new configuration each successfully transmitted photon passes through a number of photon routers 
$k_i$ that depends on which crystal it has been created from:
\begin{equation}
\label{stages}
k_i=\begin{cases} i, & \mbox{if }~  i\leqslant m-1 \\ m-1, & \mbox{if }~  i=m. \end{cases}
\end{equation}
Since each channel is subjected to a different attenuation,
we again compensate the different absorption rates by choosing $\mu_i$
(the mean number  of generated pair  of  the $i$-th crystal) as:
\begin{equation}
\label{balance}
\mu_i=\frac{\bar{\mu}}{\gamma^{k_i}}\,,\qquad i=1,\dots,m.
\end{equation}
This choice is still suboptimal but is sufficient to outperform the SMHPS in many situation of experimental interest.
To further improve the performances, an optimization over the different $\mu_i$ should be performed.
{As an example, we show in the appending that in the ideal case case of $\eta=\gamma=1$
the one photon probability can be improved by using optimized pump parameters. 
However, the optimization with imperfect efficiency and transmission
cannot be performed analytically.
}

This multiplexing architecture is therefore asymmetric and will be denoted with  AMHPS, to highlight the differences with the SMHPS scheme.
It is worth to note that both the symmetric than the asymmetric architecture
with the same number $m$ of crystals, require the same number of detectors and routers.
With each binary switch configured as before,
if two or more channels are heralded, the one that needs to pass through less routers is selected and  
routed to the end of the chain, thereby decreasing the probability of absorption. Again,
if no detector fires, the first channel is routed to the end. 
Moreover, the different delay lines should be carefully adjusted such that
each source would produce a final output photon at the same time.

As shown in  appendix \ref{sec:SMHPScalculation}, 
the probability of emitting $n$-photons  for the AMHPS is given by:
\begin{align}
\label{pnasym}
&\prn^A=\frac{[(1-\eta )\bar\mu]^ne^{-(1-\eta )\bar\mu }}{n!}
e^{-\eta\bar{\mu}\frac{(2-\gamma)\gamma^{1-m}-1}{1-\gamma}}+
\\
\notag&+\frac{\bar{\mu}^ne^{-\bar{\mu} }}{n!}
\sum_{i=1}^m e^{-\eta\bar{\mu} \frac{\gamma^{1-i}-1}{1-\gamma}}
\left[1-(1-\eta)^n e^{\eta\bar\mu}e^{-\frac{\eta\bar{\mu}}{\gamma^{k_i}}}\right].
\end{align}
It worth noting that, by using the compensation proposed in $(\ref{balance})$, the dependence of the one photon probability on the intensities is reduced to a single variable, i.e. $\bar{\mu}$.
Moreover, when $\gamma\rightarrow 1$, the value of $\prn^A$ for the asymmetric scheme coincides with the values 
$\prn^S$ (eq. \eqref{symprob})
of the symmetric scheme  for any $\eta$. Again, in the limiting case of 
 $\eta\rightarrow0$, we obtain the faint laser source with mean photon number $\bar\mu$.

In the next section we are going to confront the FL, the SMHPS and the AMHPS.

\section{Performance Comparison}
\label{sec:comparison}
This section is   devoted to the   comparison of the FL, the SMHPS and the   AMHPS by means of numerical considerations. 
We will use the performance index previously defined, namely
the maximum of the one photon probability ${\pr}_1$ provided that the SNR has a greater or equal value than $ \Theta$:
\begin{equation}
\overline{\pr}_1(\eta,\gamma, \Theta)
=\max_{\mu,\text{SNR}\geqslant  \Theta}\prnone(\mu,\eta, \gamma).
\end{equation}
Similarly to the ideal case, since $\lim_{\mu\rightarrow0}\text{SNR}=+\infty$ and $\lim_{\mu\rightarrow+\infty}\text{SNR}=0$ for both
the symmetric than the asymmetric scheme, by choosing the appropriate value of $\mu$, any value of the SNR can be achieved.
Notice that, once the number of crystals and the SNR threshold are fixed, the above quantity depends only on  $(\eta,\gamma)$.

\begin{figure}[t]%
\centering
\includegraphics[width=9cm]{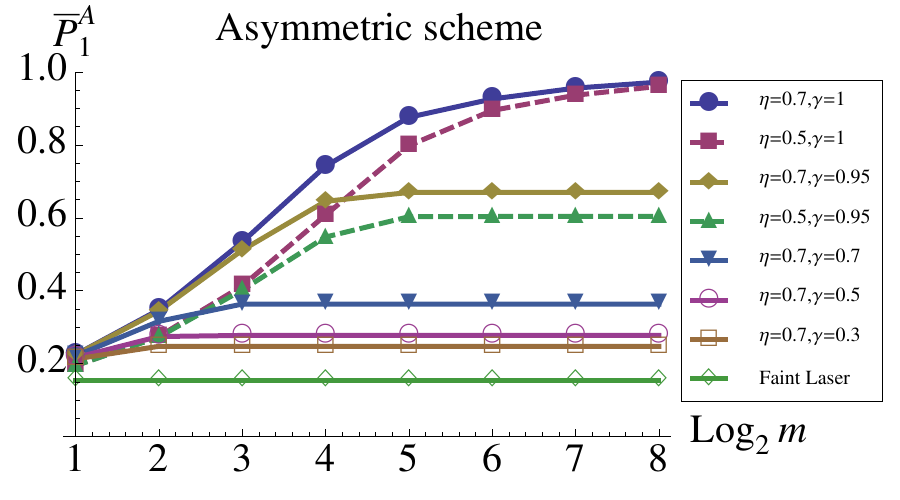}
\caption{One-photon probability  for the AMHPS    with $ \Theta=10$ and $m\in \{2,\dots, 256\}$ and various pairs of $(\eta,\gamma)$.
We also report the corresponding $\overline\pr_1=0.155$ of the faint laser.}
\label{scale}
\includegraphics[width=9cm]{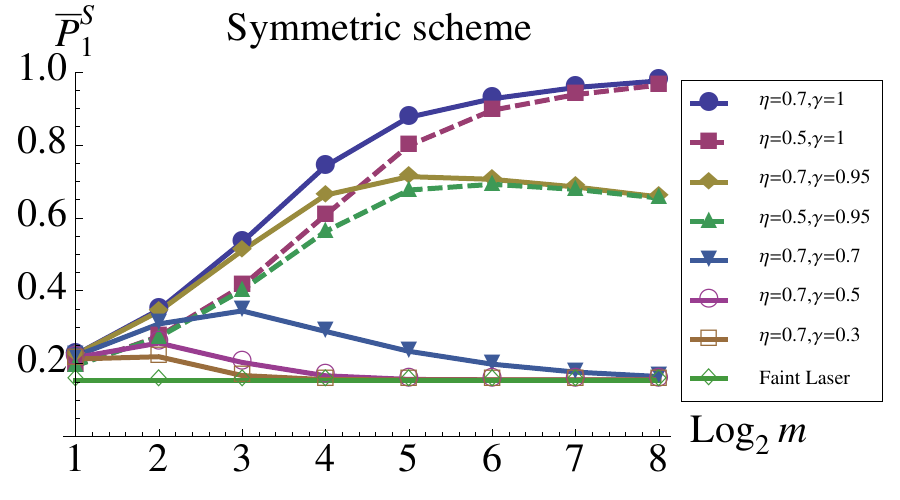}
\caption{One-photon probability  for the SMHPS    with $ \Theta=10$ and $m\in \{2,\dots, 256\}$ and various pairs of $(\eta,\gamma)$.
We also report the corresponding $\overline\pr_1=0.155$ of the faint laser.}
\label{scale1}
\end{figure}
\subsection{Scalability of the schemes for finite efficiencies}
 We first discuss  the scalability  features of the strategies with respect to the total number of crystals. 
In fig. \ref{scale}-\ref{scale1}  we plotted
 the values  of $\overline{\pr}^A_1$ and $\overline{\pr}^S_1$ versus the number of crystals, $m$, ranging from $2$ to $256$ for different pairs of $(\eta,\gamma)$ and for threshold SNR given by $\Theta=10$.

For what concern the AMHPS  it is apparent that, for all the considered pairs of $(\eta,\gamma)$, the one photon probability increases until   it reaches an asymptotic value  (the dependance of this value on $\gamma$ and $\eta$  is non-trivial). This fact implies that, once the detection efficiency and the transmissivity are fixed, there is a threshold value on the number of crystals above which there is no further improvement in the performances of the scheme. It is worth noting that for experimental realistic parameters, i.e. $\gamma\lesssim 0.5$, the asymptotic performances are practically  already reached for $m=8$.

For what it concerns the SMHPS, after an initial transient, 
the one photon probability starts to decrease (a part for the ideal case of $\gamma=1$). 
We have {analytically shown} that, excluding the $\gamma=1$ case, 
in the limit of infinite number of crystals ($m\rightarrow+\infty$) the one photon probability of the symmetric scheme
approach the one photon probability of the faint laser {when the pump parameters are chosen in order to
have asymptotically the same SNR} (as can be sees in fig. \ref{scale1}). 
{The result is demonstrated in  appendix \ref{sec:k=inf}}.
This property implies that the performance of symmetric architecture doesn't not always improve if we increase the number $m$ of crystals 
and a ``fine tuning'' of $m$ should be used in function of $\eta$ and $\gamma$ to optimize $\overline\pr^S_1$. 
Note that only if the parameters $\eta$ and $\gamma$ are perfectly know the optimization on the number of crystals can be performed
exactly. 
The asymmetric scheme, on the other side, is always improving when the number of crystals is increased:
from this point of view it is more "robust" than the symmetric scheme, since it does not require the precise knowledge of
$\eta$ and $\gamma$.

Let us try to give a motivation for this counter intuitive behavior: has we have mention above, in the SMHPS each successfully transmitted photon have to pass through $\log_2 m$ routers, increase the number of crystal means also to increase the absorption rate the photons are subjected to. Thus for this geometry architecture  the benefits deriving  from the increase of the number of crystals do not compensate the increase in the absorption rate. On the contrary, as we have mention above, in the AMHPS are most likely to be selected those channels whose photons have to pass through less routers in order to reach the global output   leading,   on average, to a lower absorption rate.

  Summarizing, there are some benefits for the SMHPS in increasing the number of crystals but only up to a certain number, depending on  the detection efficiency and transmissivity. Anyway increasing further the number of crystals will lead to  poorer and poorer performances.
  
  The AMHPS   offers significant benefits in increase the number of crystal until a certain number depending on  the detection efficiency and transmissivity, once the threshold is reached increasing further the number of crystal will left the performances unchanged.
Finally, we remark that since the two methods adopt the same post-selection rules, the gap in the performances 
arises form the different architecture geometries that is responsible  for the different distribution of the routers. 

Before closing this section, let us briefly review and discuss the scalability analysis proposed in \cite{PhysRevA.66.053805}. 
In that paper, in order to evaluate the advantages of a scheme with respect to the FL, 
the {\itshape gain}, namely the ratio $G$ between the one-photon probability of having one photon in the 
output of the SMHPS and the probability of producing one photon with the FL, is considered. 
It is worth remarking that in \cite{PhysRevA.66.053805} 
the one-photon probability for the proposed scheme is computed neglecting both detection and transmission inefficiencies: 
{in this ideal case, it turns out that both
the scheme and the FL have the same SNR provided
that the intensity with which the HS units are fed is
equal to the intensity of the FL.} 
In order to analyze the scalability taking into account the absorption due to the routing chain, they propose to consider the asymptotic behavior of the product $\gamma^k G,$ thus comparing the benefits of a growing multi-crystal architecture to the increase in the absorption rate due to the longer routing chain. As a result of this analysis, we have that the advantage of the SMHPS is maintained (i.e. $\lim_{k\rightarrow \infty}\gamma^kG>1$) if $\gamma \geq 1/2$.

   \begin{figure*}[t]
{\includegraphics[width=8cm]{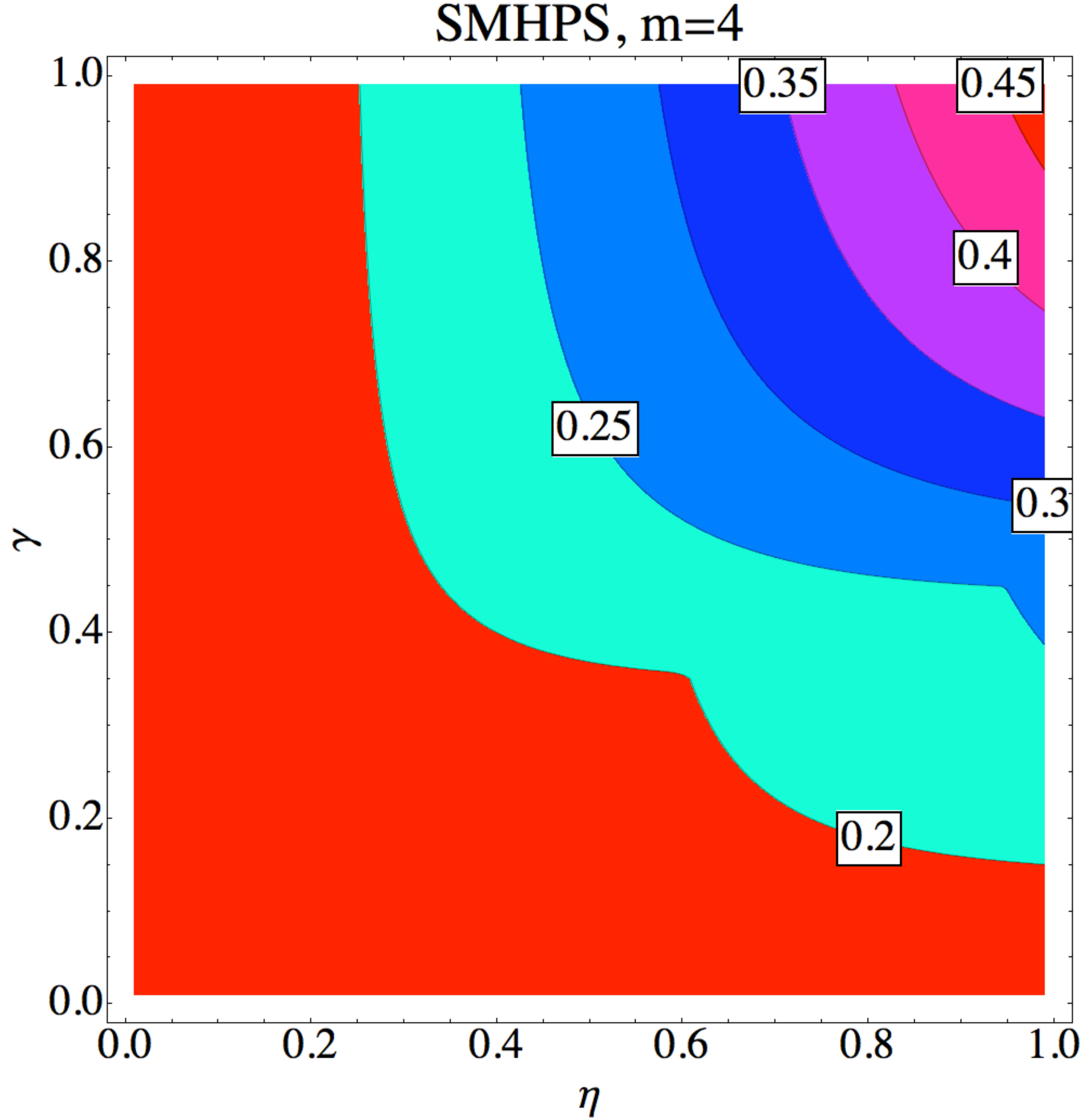}}
{\includegraphics[width=8cm]{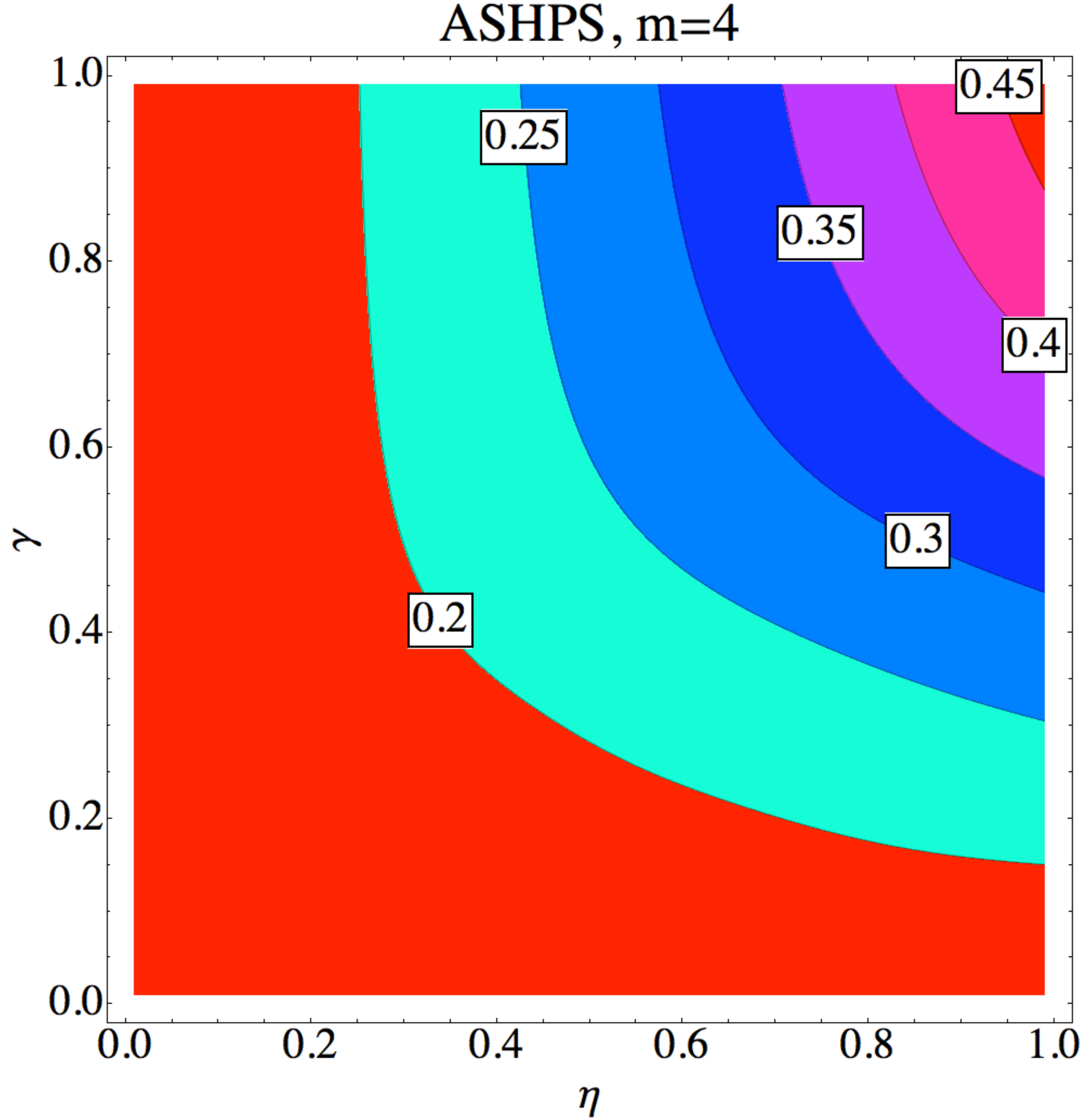}}
\caption{Contours of the one-photon probability for the SMHPS (left) and the AMHPS  (right)  with guaranteed SNR, $ \Theta=10$, and $m=4$.
For the symmetric architecture, for each $(\eta,\gamma)$ we choose the number of crystals $m'\leq m$ that maximize $\overline\pr_1$.
$\overline{\pr}_1$ is always above the value of the one-photon probability (given by $0.155$) of the FL with guaranteed SNR equal to 10.
In the SMHPS, the contour lines are not always smooth due to the changes in the $m'$ value.}
\label{m4_10}
\end{figure*}

  \begin{figure*}%
\centering
{\includegraphics[width=8cm]{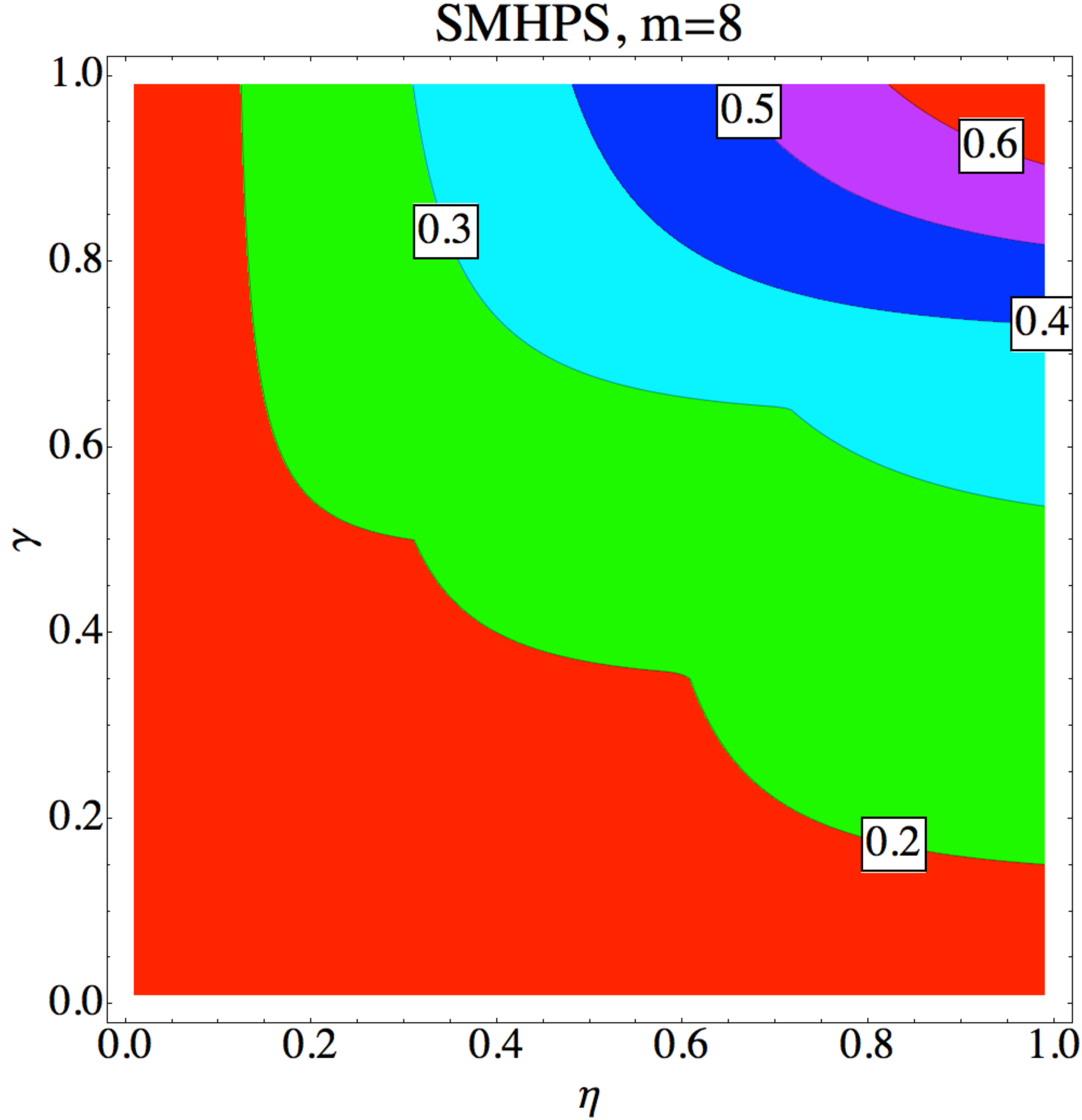}}
{\includegraphics[width=8cm]{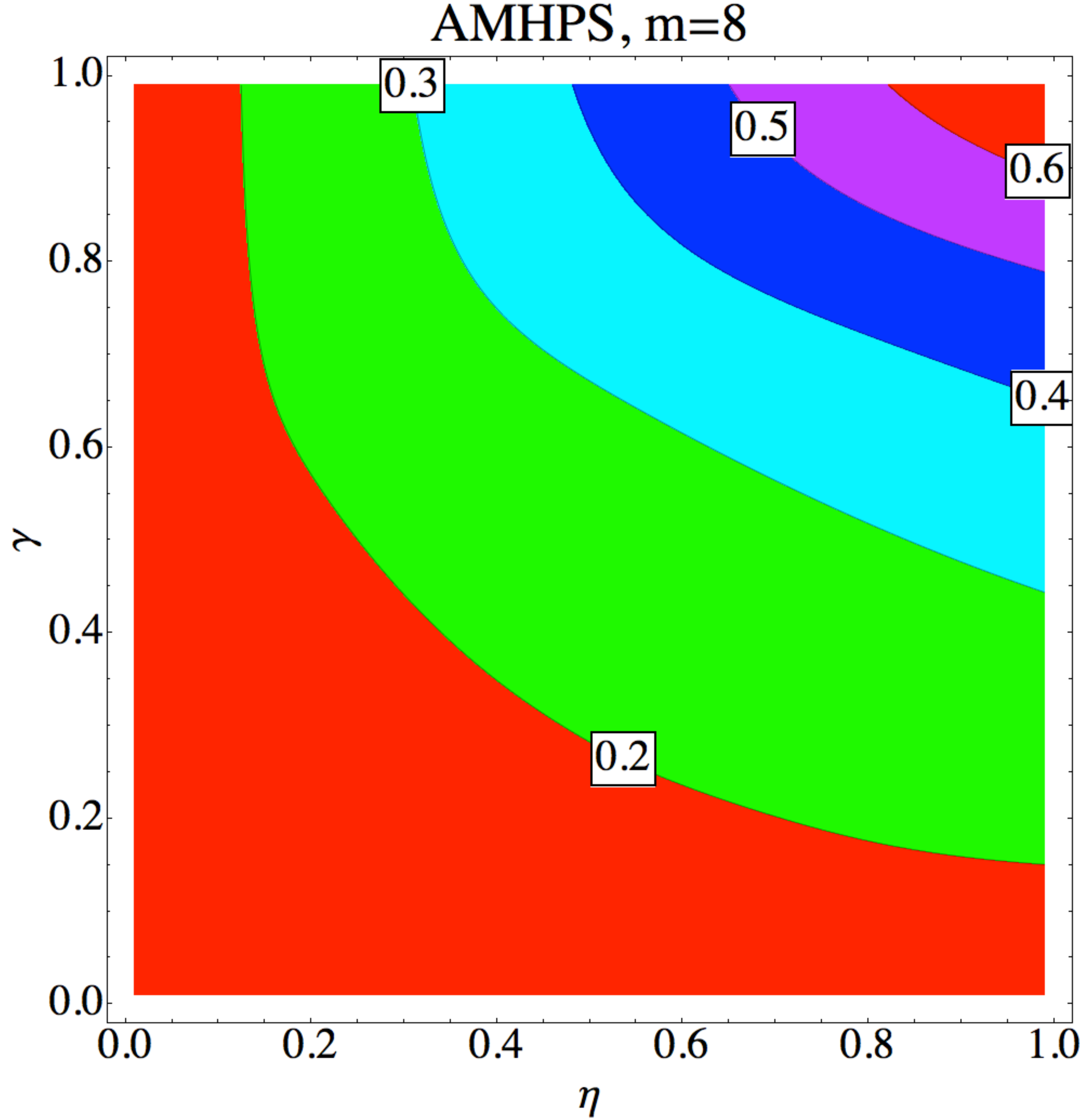}}
\caption{
Contours of the one-photon probability for the SMHPS  (left) and the AMHPS (right) with guaranteed SNR, $ \Theta=10$, and $m=8$. 
For the symmetric architecture, for each $(\eta,\gamma)$ we choose the number of crystals $m'\leq m$ that maximize $\overline\pr_1$.
$\overline{\pr}_1$ is always above the value of the one-photon probability (given by $0.155$) of the FL with guaranteed SNR equal to 10.
In the SMHPS, the contour lines are not always smooth due to the changes in the $m'$ value.
}
\label{m8_10}
\end{figure*}

In order to perform a similar analysis we should compare the asymptotic behavior, in the limit of an infinite number of crystals, 
 of the rate between the  one-photon probability for the SMHPS, $(\ref{symprob})$, with the probability of producing one photon with the FL, 
{for the same SNR. As shown in appendix \ref{sec:k=inf}, in the $k\rightarrow\infty$ limit the SNR of the faint laser with
intensity $\mu$ is equal to the SNR of the SMHPS with  $\mut=\mu$ (the number of crystal is $m=2^k$).}
Moreover, as explained previously, in order to avoid infinite power we need to rescale the pump power {(for both the FL than the SMHPS)}
 with the number of crystal,
$\mu\rightarrow\mu/2^k$. The gain we obtain is given by:

 \beq
\widetilde G= \frac{1-e^{-\frac{1-\gamma^k }{(2\gamma) ^k}\eta  \mu}(1-\eta )-e^{-\frac{\eta  \mu }{\gamma ^k} }
[1-e^{\frac{\eta  \mu}{2^k} }(1-\eta )]}{1-e^{-\frac{\eta  \mu }{(2\gamma) ^k}}}
 \eeq
 
In the infinite $k$ limit, the gain $\widetilde G$ is {greater than 1} 
for $\gamma\geq\frac12$ (and is actually divergent for $\gamma>1/2$) as found in 
\cite{PhysRevA.66.053805}.
 However, in the rescaled pump power case, the one-photon probability of the SMHPS tends asymptotically to zero and 
 the gain does not seems the proper index to evaluate the absolute performance of a scheme. In fact, despite exhibiting an advantage with respect to the FL,  $\pr_1$ is asymptotically zero in both cases (as already said, without the 
 rescaling $\mu\rightarrow\mu/2^k$, the single photon probability of the SMHPS and of the FL coincide in the large $k$ limit, 
 and the gain is always 1). 
 This means that, with finite transmissivity, not only increasing the number of crystal beyond a certain value does not bring any advantage, but it is actually detrimental to the SMHPS scheme performance. On the other hand, the AMHPS scheme is ``robust'' with respect to implementations with large numbers of crystals, as it is clearly shown in fig. \ref{scale}.

\subsection{Comparison between AMHPS and SMHPS}

   \begin{figure*}[t]
\centering
\includegraphics[width=8.6cm]{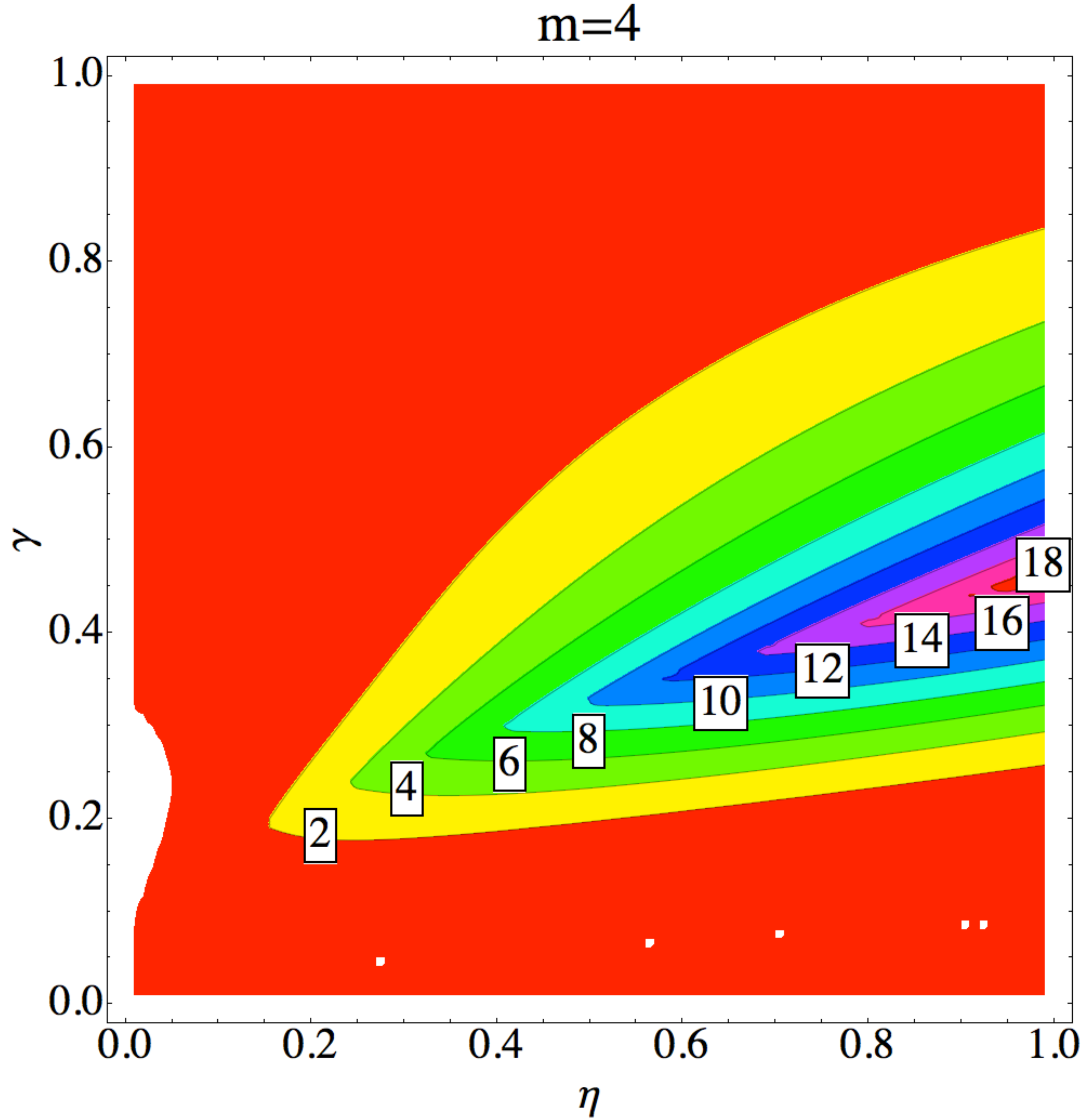}
\includegraphics[width=8.6cm]{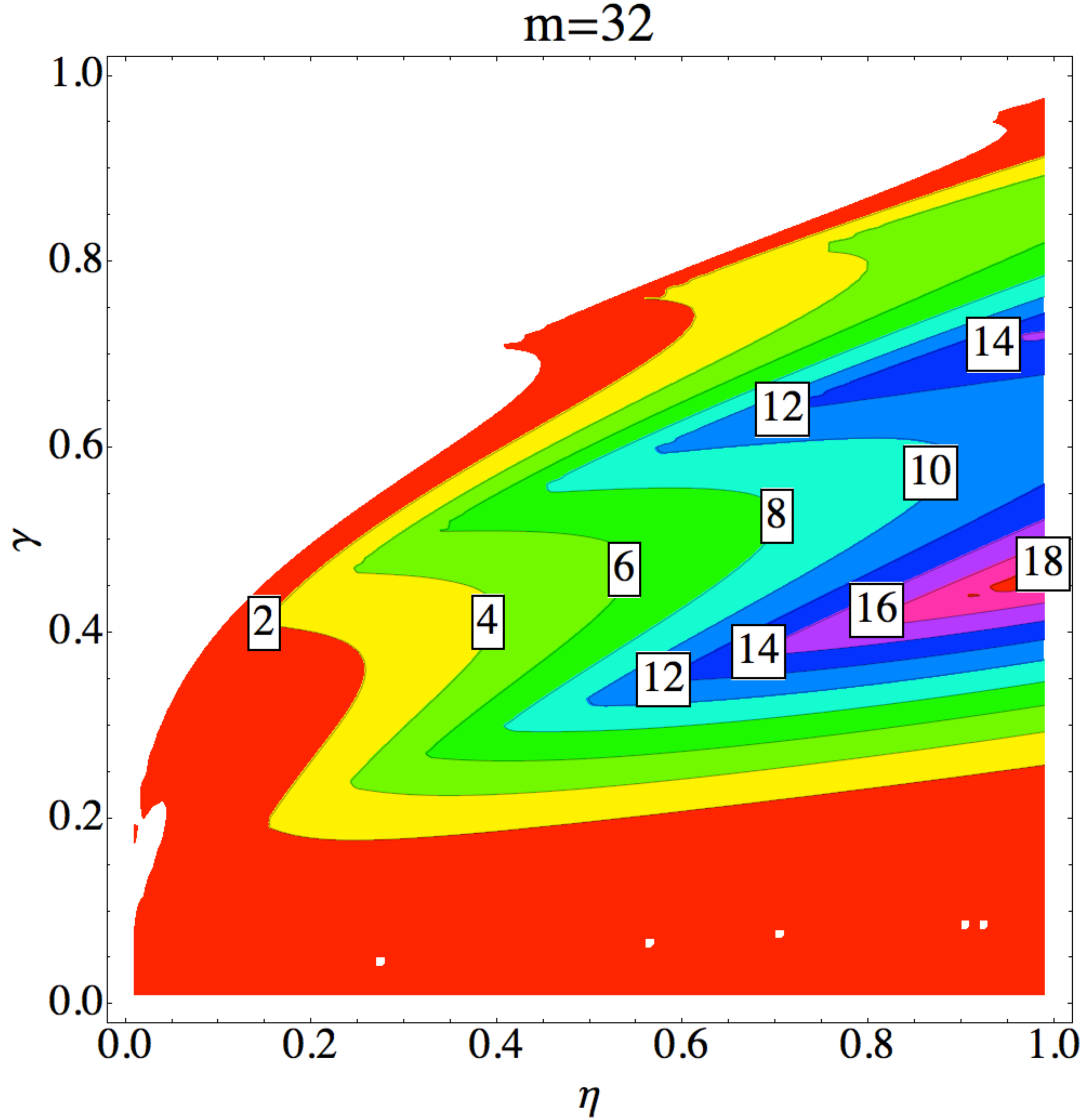}
\caption{Contours of the percentage differences \eqref{delt} with guaranteed SNR, $ \Theta=10$, and $m=4$ (left) or $m=32$ (right). 
Again, for the symmetric architecture, for each $(\eta,\gamma)$ we choose the number of crystals $m'\leq m$ that maximize $\overline\pr_1$.
In the white area the SMHPS is performing better than the AMHPS. }
\label{m4_8}
\end{figure*}

We here compare the performances of the AMHPS and the SMHPS by fixing the guaranteed SNR and the number of crystals $m$.
However, in order to obtain a fair comparison we analyze the asymmetric scheme with $m$ crystals with the
symmetric scheme with $m'\leq m$ crystals: in fact, as we have previously shown, the performance of the symmetric scheme does not
always improve when the number of crystals is increased. Thus, for each $(\eta,\gamma)$,
the asymmetric scheme with $m$ crystals must be compared with
the symmetric scheme with $m'$ crystals, where $m'\leq m$ is chosen in order to maximize $\overline\pr_1$.
In Figures \ref{m4_10} and \ref{m8_10} 
we displayed the contours plot of  $\overline{\pr}_1$ for the AMHPS and the SMHPS 
with SNR threshold given by $ \Theta=10$ and $m=4$ and $8$ respectively. 
When  $\Theta=10$ the one photon probability $\overline{\pr}^{FL}_1$ of the faint laser is given by $0.155$.
We can see   that both the AMHPS and the SMHPS always outperform the FL in the plane $(\eta,\gamma)$. 
As expected, the best performances are reached for high values of both the detection efficiency and the transmissivity: furthermore  
in this limit the two methods are comparable
 since for $\eta\rightarrow1$ and $\gamma\rightarrow1$
they both tend to the ideal situation of the MHPS. 
 
We can also define:
\begin{equation}
\label{delt}
\Delta(\eta,\gamma,\Theta):=100\frac{\overline{\pr}^A_1-\overline{\pr}^S_1}{\overline{\pr}^S_1}
\end{equation}
as the percentage differences between the two optimized single photon probabilities.
In Fig. \ref{m4_8} we shown the contour of $\Delta$ for an SNR equal to $10$ and $m=4$ and $32$. 
The AMHPS  outperforms the SMHPS in a vast portion of the plane $(\eta,\gamma)$. Anyway, the advantage of the AMHPS {in more realistic situations} is apparent, especially in the {area} where $\gamma\approx 0.5$ and the detection efficiency is {higher than 0.5}, the AMHPS   outperforms the SMHPS.

\section{Experimental feasibility and Conclusions}

Let us now discuss a possible experimental realization of the proposed SPS configuration.
Nowadays, integrated devices represent the best resource to achieve high efficiency of the SPDC process and ensure good coupling into
single mode fibers (for a review on
integrated source see \cite{tanz12lpr}). 
It is possible to use non-degenerate collinear phase matching and a dichroic beam splitter (or alternatively using a counterpropagating mode source \cite{cail10ope}) in order to separate the two beams.
For instance, it was recently reported the possibility of heralding single telecom photons at 4.4 MHz rate with $45\%$ heralding efficiency
\cite{poma12qph}. Moreover, to efficiently detect the triggered photon, 
high efficient transition-edge sensors (TES) can be used: an heralding efficiency of $\eta\sim62\%$
has been recently reported by using TES \cite{smit12nco},
while 810nm single photon heralded source with $83\%$ heralding efficiency has been shown in \cite{rame12qph}
 (see also \cite{eisa11rsi} for a review on single photon detectors).
Regarding the optical switch, a 2x2  silicon  electro-optic  switches  with a broad bandwidth (60 nm), an ultrafast speed (6 ns) 
and a transmission of $\gamma\sim50\%$ has been reported \cite{dong10ope}; other modulators otherwise allow 
lower losses at the cost of reduced working spectrum \cite{reed10npho}.
As shown in Fig. \ref{m4_10}, with these values of $\eta$ and $\gamma$,
the asymmetric scheme is more performant than the symmetric one.

In conclusion we have proposed an asymmetric architecture for the multiplexed heralded single photon source and we
have compared it with the symmetric version proposed in \cite{Shapiro:07} and with the faint laser source by using a
performance index $\overline\pr_1$ we introduced.
We have proven that the asymmetric architecture outperform the symmetric scheme
in a vast region of the parameter space $(\eta,\gamma)$ and both outperform the FL for any values of $\gamma$ and $\eta$.
We have also {demonstrated} that, in the large number of crystal limit and by considering a fixed SNR, 
the symmetric configuration is asymptotic{ally equivalent} to the faint laser for any $\gamma\neq1$, while for the asymmetric scheme
the one photon probability increases until it reaches an asymptotic value dependent on $\gamma$ and $\eta$. 
This implies that the symmetric architecture requires a ``fine tuning'' of $m$ 
in function of $\eta$ and $\gamma$ to optimize $\overline\pr^S_1$.
On the other side, when the number of crystals is increased, 
the asymmetric architecture is always improving its performances.
Values of $\overline\pr^A_1$ close to the asymptotic ones, at least for experimentally available efficiencies, are reached already around 8 crystals. This implies that, even if the expected energetic consumption for the AMHPS scheme is higher due to the implemented pre-compensation for the losses in the routing chain, the necessary overhead will be very limited.
We believe that our results will be relevant {to} any future realization of heralded single photon source based on multiplexed architecture.

\begin{acknowledgments} 
The Authors acknowledge the Strategic-Research-Project QUINTET of the Department of
Information Engineering, University of Padova and the
Strategic-Research-Project QUANTUMFUTURE of the University of
Padova.
\end{acknowledgments}

 \appendix

 \section{Statistics of the heralded sources}
\label{sec:SMHPScalculation}
We here provide a derivation of the MHPS statistics for the symmetric and asymmetric architecture
with the number of crystals given by $m$. 
If we denote by $\mut_i$ the mean number of generated pairs from the $i$-th crystal and with $k_i$
the number of routers that the signal photon generated by the $i-$th source needs to pass through, the 
asymmetric and symmetric architectures only differ from the expression of $k_i$ and $\mut_i$:
in the AMHPS, $k_i$ is given by \eqref{stages} and $\mut_i=\bar\mu/\gamma^{k_i}$, while for the SMHPS we have $k_i=k\equiv\log_2m$
and $\mut_i=\mut/\gamma^k$, $\forall i$ .
We thus calculate the statistic of the output in this general framework.

The probability that the source $i$ doesn't trigger is given by $p_{i}=e^{-\eta \mut_l}$.
Let's denote by $\chi$ the first HS, starting from $i=1$, that triggers. If no source triggers we set $\chi=0$.
The probability that $\chi=i$ is given by: 
   \begin{align}
\notag\pr(\chi=i)&=\begin{cases} (1-p_i)\prod_{\ell=1}^{i-1}p_\ell, & \mbox{if}~ i\neq 0, \\\prod_{\ell=1}^{m}p_\ell &  \mbox{if}~ i=0. \end{cases}
\\
&=\begin{cases} (1-e^{-\eta \mut_i})e^{-\eta \sum_{\ell=1}^{i-1}\mut_\ell}, & \mbox{if}~ i\neq 0, \\
e^{-\eta \sum_{\ell=1}^{m}\mut_\ell} &  \mbox{if}~ i=0. 
\end{cases}
\end{align}
  The probability of having $j$ photons in the $i$-th signal channel (before the switches), provided that  $\chi=i\neq 0$ is
\begin{equation}
\label{ }
\pr(N_i=j|\chi=i)=
\frac{\frac{\mut_i^j}{j!} e^{-\mut_i}(1-(1-\eta)^j) }{(1-e^{-\eta \mut_i})}, \quad \mbox{if}~ i\neq 0 
\end{equation}
where $N_i$ is the number of photons generated at the i-th source.
When $\chi=0$ (no source triggers), the router will select the first source and the probability of
having $j$ photon on channel 1, provided that no sources have triggered is
\begin{equation}
\label{ }
\pr(N_1=j|\chi=0)=
\frac{\frac{\mut_i^j}{j!} e^{-\mut_i}(1-\eta)^j }{e^{-\eta \mut_i}}
\end{equation}

The probability of having $n$ photons in the final output provided that $\chi=i\neq0$ and $N_i=j$ is given by:
\begin{align}
&\pr(\cn=n|N_i=j,\chi=i)=
\\
\notag
&=\begin{cases} 
{j \choose n}(\gamma^{k_i})^n(1-\gamma^{k_i})^{j-n}, & \mbox{if}~ i\neq 0~\mbox{and}~n\leqslant j
\\
0 & \mbox{if}~ i\neq 0~\mbox{and}~n> j
\end{cases}
\end{align}
while
$\pr(\cn=n|N_1=j,\chi=0)=
{j \choose n}(\gamma^{k_1})^n(1-\gamma^{k_1})^{j-n},$ if $n\leqslant j$.
In the previous expression $\hat N$ is the number of photons generated at the global output.

 Finally the probability of having $n$ photons in the final output is given by:
\begin{widetext}
\begin{align}
\label{pn_general}
\prn=
&\sum_{j=n}^{\infty}\pr(\cn=n|N_1=j,\chi=0)\times\pr(N_1=j|\chi=0)\pr(\chi=0)+
\\\notag
&+\sum_{i=1}^{m}\sum_{j=n}^{\infty}\pr(\cn=n|N_i=j,\chi=i)\pr(N_i=j|\chi=i)\pr(\chi=i)
\\
\notag&=\frac{[\mut_1\gamma^{k_1}(1-\eta)]^ne^{-\mut_1\gamma^{k_1}(1-\eta)}}{n!}e^{-\eta \sum_{\ell=1}^{m}\mut_\ell}
+\sum_{i=1}^m 
\frac{(\mut_i \gamma^{k_i})^n e^{-\mut_i \gamma^{k_i}} }{n!}\left[1-(1-\eta)^n e^{-\eta(1-\gamma^{k_i})\mut_i}\right]
e^{-\sum_{\ell=1}^{i-1}\eta \mut_\ell}.
\end{align}
\end{widetext}
Let's now specialize the last result for the symmetric case with $\mut_i=\bar\mu/\gamma^k$ and $k_i=k$. We obtain
\begin{align}
\pr^S_n
&=\frac{[\mub(1-\eta)]^ne^{-\mub(1-\eta)}}{n!}e^{-\eta\mub \frac{2^k}{\gamma^k}}
+\\
&
\notag\frac{\mub^n e^{-\mub} }{n!}\left[1-(1-\eta)^n e^{-\eta\mub(\frac{1}{\gamma^{k}}-1)}\right]
\sum_{i=1}^m e^{-(i-1)\eta \frac{\mub}{\gamma^k}}.
\end{align}
and performing the last sum we obtain \eqref{symprob}.

In the asymmetric case we have $\mut_i=\mut/\gamma^{k_i}$ and the $k_i$ are given by \eqref{stages}. We obtain
\begin{align}
\pr^A_n
&=\frac{[\mut(1-\eta)]^ne^{-\mut(1-\eta)}}{n!}e^{-\eta\mut(\frac{1}{\gamma^{m-1}}+\sum^{m-1}_{\ell=1} \frac{1}{\gamma^\ell})}
+\\
&
\notag\frac{\mut^n e^{-\mut} }{n!}
\sum_{i=1}^m \left[1-(1-\eta)^n 
e^{-\eta\mut(\frac{1}{\gamma^{k_i}}-1)}\right]
e^{-\eta \mut_\ell\sum_{\ell=1}^{i-1}\frac{1}{\gamma^\ell}}.
\end{align}
and performing the sum on the exponents we obtain \eqref{pnasym}.

\section{Analysis of the 2-crystal architecture: Symmetry is not optimal}
 \label{sec:sym_not_opt}
In this section we will prove that the (ideal) 2-crystal architecture where the two crystals are driven with same pump laser intensity
represents a suboptimal choice for the one photon probability.
 We here consider a MHPS composed by two crystals each one  fed 
 with different intensities such that the mean number of emitted pairs are $\mu_1$ and $\mu_2$ respectively. 
The probability of having $n$ photons in the output is obtained from equation \eqref{pn_general} by using
$m=2$, $\gamma=1$ and $\eta=1$:
\beq
\pr_n=\delta_ne^{-\mu_1-\mu_2}+(1-\delta_n)(\frac{\mu^n_1e^{-\mu_1}}{n!}+\frac{\mu^n_2e^{-\mu_1-\mu_2}}{n!})
\eeq

\begin{figure}[t]
\centering
\includegraphics[width=8.8cm]{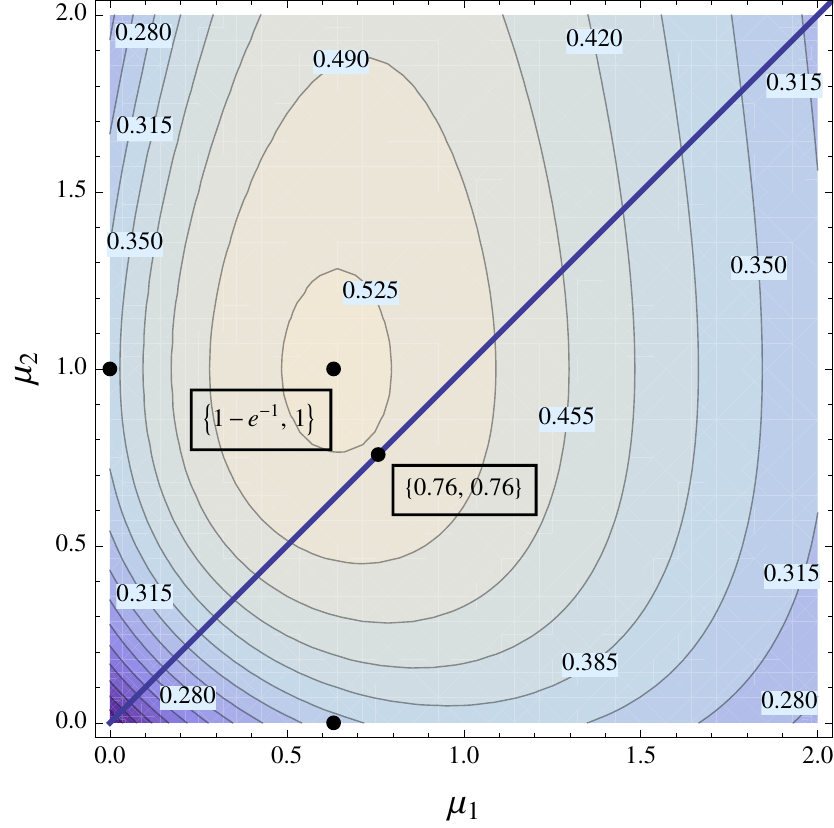}
\caption{Contours of the one-photon probability $(\ref{oneid})$ for the two crystals MHPS. The blue line represent the  $\mu_1=\mu_2$ case.}
\label{asymax}
\end{figure}

In particular, the probability of having one photon in the final output is given by:
\begin{equation}
\label{oneid}
\prone=\mu_1e^{-\mu_1}+\mu_2 e^{-(\mu_1+\mu_2)}
\end{equation}
and its maximum is achieved when
\begin{equation}
\label{ }
(\mu_1,\mu_2)=(1-e^{-1},1).
\end{equation}
This can be seen in fig. \ref{asymax} where the contours  of $\prone$ are plotted: the blue line represent the cases $\mu_2=\mu_1$.
Also by using the performance parameter introduced in section \ref{performance} it is straightforward to show that
by using $\mu_1\neq\mu_2$ leads to better performances with respect to the choice $\mu_1=\mu_2$.

Therefore, we   conclude that, in general, feeding all the crystals with the same laser intensities,  leads to  suboptimal performances.
It is worth noticing that the asymmetry comes from the switch selection: the
first source is initially checked and, only if this source doesn't trigger, the switch considers the second source. 
As we demonstrated, this asymmetry can be exploited to improve the probability of having one photon at the output channel.
 
 In the ideal case (perfect detection and transmission), it is straightforward to compute the intensities that lead to the one photon probability optimal value generalizing the analysis above. Anyway, such optimization task become quite non-trivial if finite efficiency are taken into account, especially for a large number of crystals.

 \section{Infinite crystals limit of the SMHPS}
\label{sec:k=inf}
{{We here} analytically show that the one photon probability of the
 symmetric scheme approach{es} the one photon probability of the faint laser
when the pump {intensities} are chosen in order to have the same SNR.
The SNR of the SHMPS  is
\begin{widetext}
\beq
\label{SNRsymeta=1}
\frac{\mut \left[1-e^{ -\frac{\eta  \mut}{\gamma ^{k}}  }e^{ \eta  \mut } (1-\eta )
-e^{-\frac{2^k}{\gamma ^{k}} \eta  \mut }\left(1-e^{ \eta  \mut } (1-\eta )\right)\right]}
{e^{\mut }-1-\mut -e^{-\frac{\eta  \mut }{\gamma ^k}}\left[(e^{\mut }-e^{ \eta  \mut }(1+\mut -\eta  \mut )\right]+e^{-\frac{2^k}{\gamma ^k} \eta  \mut } \left[(1+\mut )-e^{ \eta  \mut } (1+\mut -\eta  \mut )\right]}
\eeq
\end{widetext}

In the large $k$ limit, when $\gamma\neq1$, the previous expression is equal to the SNR of the faint laser, namely 
$\frac{\mut}{e^{\mut }- 1-\mut}$.
When $\mut=\mu$, the SNRs are equal in the large $k$ limit and
the one photon probability of the SHMPS becomes
\begin{align}
\notag
\pr^S_1&=
\mu  e^{-\mu }\frac{1-e^{-\frac{\eta  \mu }{\gamma ^k}}e^{ \eta  \mu }(1-\eta )-e^{-\frac{2^k}{\gamma ^k} \eta  \mu }[1-e^{\eta  \mu }(1-\eta )]}{1-e^{-\frac{\eta  \mu }{\gamma ^k}}}
\\
&\sim\mu e^{-\mu } \quad\text{ for }k\rightarrow\infty\text{ and }\gamma\neq1\,,
\end{align}
asymptotic, in the large number of crystal limit, to the one photon probability of the faint laser.}


\end{document}